\documentclass[fleqn,twoside]{article}
\usepackage{espcrc2}
\usepackage{epsfig}

\usepackage{graphicx}
\usepackage[figuresright]{rotating}

\newcommand{\AmS}{{\protect\the\textfont2
  A\kern-.1667em\lower.5ex\hbox{M}\kern-.125emS}}

\title{Issues in the Extraction of $m_s$ and $V_{us}$ from Hadronic
$\tau$ Decay Data}

\author{K. Maltman\address{Department of Math and Stats, York Univ., 
Toronto, ON, M3J 1P3, Canada and
CSSM, Univ. of Adelaide, Adelaide, SA, 5005, Australia}
\thanks{e-mail: kmaltman@yorku.ca. Work supported by a grant from
the Natural Sciences and Engineering Research Council of Canada}}
       
\begin{document}

\begin{abstract}
Various complications encountered in the process of attempting 
to extract the basic Standard Model parameters, $m_s$ and $V_{us}$, 
from hadronic $\tau$ decay data are discussed. 
\vspace{1pc}
\end{abstract}

\maketitle

\section{Background}
Hadronic $\tau$ decays into states with zero (non-zero) net strangeness
provide access to the spectral functions of the correlators of
the flavor $ij=ud$ ($us$) vector (V) and axial vector (A)
currents, $J^\mu_{V/A;ij}$. Explicitly, defining
$\Pi^{(J)}_{V/A}$, the $J=0,1$ parts of a given V/A correlator,
and $R_{V/A;ij}$ by
\begin{eqnarray}
&&i\, \int\, d^4x\, e^{iq\cdot x}\langle 0\vert T\left( J_{V/A}^\mu (x)
J_{V/A}^\nu (0)\right)\vert 0\rangle =\nonumber\\
&&\quad \left( q^\mu q^\nu -q^2g^{\mu\nu}\right)\Pi^{(1)}_{V/A}(q^2)
+q^\mu q^\nu \Pi^{(0)}_{V/A}\ ,\\
&&R_{V/A;ij}
\equiv {\frac{\Gamma [\tau^- \rightarrow \nu_\tau
\, {\rm hadrons}_{V/A;ij}\, (\gamma)]}
{\Gamma [\tau^- \rightarrow
\nu_\tau e^- {\bar \nu}_e (\gamma)]}}\ ,
\end{eqnarray}
with ($\gamma $) denoting extra photons and/or lepton pairs,
$R_{V/A;ij}$ can be expressed in terms of a
weighted integral involving the corresponding spectral functions, 
$\rho^{(J)}_{V/A;ij}(s)$~\cite{braatenetc,pichrev}:
\begin{eqnarray}
&&{\frac{R_{V/A;ij}}{\left[ 12\pi^2\vert V_{ij}\vert^2 S_{EW}\right]}}
=\int^{1}_0\, dy_\tau \,
\left( 1-y_\tau\right)^2 \nonumber\\
&&\qquad\left[ \left( 1 + 2y_\tau\right)
\rho_{V/A;ij}^{(0+1)}(s) - 2y_\tau \rho_{V/A;ij}^{(0)}(s) \right]\nonumber \\
&&\qquad\equiv \int_0^{m_\tau^2}ds\, {\frac{dR_{V/A;ij}(s)}{ds}}\ ,
\label{taukinspectral}
\end{eqnarray}
where $y_\tau =s/m_\tau^2$, $V_{ij}$ is the
flavor $ij$ CKM matrix element, and $S_{EW}$ is 
an electroweak correction. The experimental decay distribution,
$dR_{V/A;ij}(s)/ds$, thus yields the linear combination
$w_T(y_\tau )\rho_{V/A;ij}^{(0+1)}(s)+w_L(y_\tau )\rho_{V/A;ij}^{(0)}(s)$,
where $w_T(y)\equiv (1-y)^2(1+2y)$, $w_L(y)\equiv -2y(1-y)^2$.
The $(J)=(0)$ part of 
Eq.~(\ref{taukinspectral}), and of analogous OPE expressions and/or
expressions involving different weights, will be
called ``longitudinal'' in what follows.

The combinations, $\rho_{V/A;ij}^{(0+1)}(s) \equiv \rho_{V/A;ij}^{(0)}(s)+
\rho_{V/A;ij}^{(1)}(s)$ and $s\rho^{(0)}_{V/A;ij}(s)$, 
in Eq.(\ref{taukinspectral}) correspond to correlator combinations
(generically $\Pi (s)$) having no kinematic singularities and hence
satisfying the general FESR relation
\begin{equation}
\int_0^{s_0}w(s)\, \rho(s)\, ds\, =\, -{\frac{1}{2\pi}}\oint_{\vert
s\vert =s_0}w(s)\, \Pi (s)\, ds\ ,
\label{basicfesr}
\end{equation}
valid for any $w(s)$ analytic in the region $\vert s\vert <M$
with $M>s_0$. Use of the OPE on the RHS of Eq.~(\ref{basicfesr})
allows one to determine OPE parameters 
in terms of experimental spectral 
distributions{\begin{footnote}{For ``intermediate''
scales such as those involved in hadronic $\tau$ decay, it turns out
that reliable use of the OPE in Eq.~(\protect\ref{basicfesr}) requires
the suppression of contributions from that part of the contour 
$\vert s\vert =s_0$ near the timelike point~\protect\cite{kmfesr}.
This is most easily accomplished by working with
``pinched'' weights, $w(s)$, i.e. those having
a zero at $s=s_0$.}\end{footnote}}. Many authors
have employed the ``$(k,m)$ spectral weight sum rules'', for which
the integrand on the LHS of Eq.~(\ref{basicfesr})
is $(1-y_\tau )^ky_\tau ^m \, dR_{V/A;ij}(s)/ds$.
These sum rules are ``inclusive'', in the sense that the $(k,m)$
spectral integrals, denoted $R^{(k,m)}_{V/A;ij}$,
can be constructed from the experimental distribution
without having to first perform a separation of the
$J=0$ and $J=1$ components. 

In flavor-breaking differences such as
$\Pi^{(0+1)}_{V/A;ud}(s)-\Pi^{(0+1)}_{V/A;us}(s)$ or
$s\, \Pi^{(0)}_{V/A;ud}(s)$-$s\, \Pi^{(0)}_{V/A;us}(s)$,
the leading $D=0$ OPE terms cancel, leaving as leading contribution
a $D=2$ term, essentially proportional to $m_s^2$.
Appropriately-weighted FESR's involving such differences
thus, in principle, allow one to determine $m_s$.
Flavor-breaking spectral combinations of the $(k,m)$ spectral
weight type are constructed by forming
\begin{equation}
\Delta R^{(k,m)}_{V/A;ij}\equiv {\frac{R^{(k,m)}_{V/A;ud}}{V_{ud}^2}}
- {\frac{R^{(k,m)}_{V/A;us}}{V_{us}^2}}\ .
\end{equation}
A number of determinations of $m_s$
using flavor-breaking sum rules have appeared in the
literature~\cite{mstauinclusive,kmms00,tauvus,pich040p1}. 
We discuss some non-trivial
complications, not all of which have been effectively tamed in
the majority of these analyses, below.

Recently it has been realized that $\tau$-decay-based, 
flavor-breaking sum rules also provide a novel
method for extracting $V_{us}$~\cite{tauvus},
one whose systematics are completely independent of those 
associated with alternate determinations based either on combining
lattice results for $f_\pi /f_K$ with $\pi_{\ell 2}$
and $K_{\ell 2}$ data~\cite{marciano04} or on $K_{\ell 3}$~\cite{ke3}.

The $\tau$-decay-based determination is competitive for the
following reason. The difference of rescaled $ud$ and $us$ spectral integrals
corresponds to an integrated correlator difference with
exactly cancelling $D=0$ OPE contributions only if the
correct $V_{us}$ is used to rescale the experimental $us$ data.
An incorrect $V_{us}$ leaves a residual $D=0$ contribution.
Since the $D=2$ term in the $us$ V+A OPE is
{\it very} small compared to the leading $D=0$ term, even
a small error in $V_{us}$ yields a sizeable $D=0$ residual.
The integrated $D=0$ and $D=2$ OPE contributions scale
differently with $s_0$, so such a $D=0$ residual can be detected
by studying the $s_0$ dependence of the two sides of the resulting FESR. 
If one has external information on $m_s$, this can also be used as
input on the OPE side in order to determine the $V_{us}$ value 
for which the desired $D=0$ cancellation occurs.
An ideal sum rule for this purpose would be one for which
the weighted $D=2$ OPE integrals were as small as possible, 
reducing errors associated with uncertainties in the input value of $m_s$,
and/or possible slow convergence of the integrated $D=2$ series.
For such a sum rule, the fractional uncertainty on $V_{us}$ 
would be essentially half that on the integrated $us$ spectral integral,
a figure which could be rather small when B factory spectral data
is finally available. This ideal situation is not as well realized as
one might hope for most of the weights discussed to date 
in the literature. We discuss this point further
below.

\section{Complications in the Extraction of $m_s$ and $V_{us}$
From Hadronic $\tau$ Decay Data}
Four main complications are encountered in analyzing flavor-breaking
$\tau$-decay-based FESR's: (i) the bad behavior of the 
integrated $D=2$ longitudinal OPE series; (ii) the convergence 
(order by order in $\alpha_s$) of the integrated $(J)=(0+1)$ $D=2$ OPE series; 
(iii) the role of possible $D>6$ OPE contributions 
and (iv) strong $ud$-$us$ spectral integral cancellations
for many of the weights employed in the literature
(leading to significant {\it fractional} errors on the resulting difference,
and hence also in the extracted values of $m_s$).

\subsection{The Integrated $D=2$ Longitudinal OPE Contribution}
It has been known for some time that
the convergence of the integrated $D=2$ longitudinal OPE series 
relevant to the determination of $m_s$ is very poor~\cite{longconv}. 
In ``contour improved perturbation theory'' (CIPT), e.g., even 
at the highest possible scale, $s_0=m_\tau^2$, 
the integrated $(0,0)$ series  
behaves as $1+0.78+0.78+0.90+\cdots$. The non-convergence
is even worse in fixed order perturbation
theory, and/or for $s_0<m_\tau^2$. One obvious solution is to 
restrict one's attention to the better-behaved $(J)=(0+1)$ sum rules, 
where the problem does not arise. This, however, requires 
a separation of the $J=0,1$ components of the 
experimental spectral distribution, which is not feasible
experimentally at present. Many analyses have, therefore, 
employed ``inclusive'' ($(J)=(0)$ plus $(J)=(0+1)$) sum rules
and attempted to assign conservative errors
to the $O(\alpha_s^3)$-truncated, badly-converged 
integrated $D=2$ longitudinal OPE sum.
This procedure turns out to violate inequalities
among longitudinal contributions to the flavor-breaking
$(k,0)$ spectral weight sum rules
which folow from the positivity of 
$\rho^{(0)}_{V;us}$ and/or $\rho^{(0)}_{A;us}$~\cite{kmlong}.
We elaborate on this point in the next paragraph.

While the $\pi$ and $K$ pole contributions to
the longitudinal spectral functions are well known experimentally,
the ``continuum'' contributions (beginning at $s_{th}^V=(m_K+m_\pi )^2$
and $s_{th}^A=(m_K+2m_\pi )^2$ in the $us$ V and A channels, respectively)
are not. For the flavor $ud$ correlators, these
are proportional to $m_{u,d}^2$, and numerically
negligible. The chiral suppression is of $O(m_s^2)$ and hence much
less strong for the $us$ V, A correlators. The basic FESR relation,
combined with the known pole term values,
ensures that any prescription for handling the weighted 
longitudinal $us$ $D=2$ OPE series translates into a statement about
the correspondingly-weighted longitudinal continuum spectral integral.
Denote this contribution by $\left[ \Delta^{(k,m)}\right]^c_L$
for the $(k,m)$ spectral weight case. Then, since $0< (1-y_\tau )<0.87$ for 
$s_{th}^V<s<m_\tau^2$, spectral positivity
ensures that the $\left[ \Delta^{(k,0)}\right]^c_L$ must
(i) be a decreasing function of $k$ and (ii) satisfy the rigorous inequalities
\begin{equation}
\left[\Delta^{(k+1,0)}\right]_L^c< 0.87 \left[\Delta^{(k,0)}\right]_L^c\ .
\label{rigorous}
\end{equation}
For kinematic reasons, one expects the $K(1460)$ and $K_0^*(1430)$ resonances
to dominate $\left[ \Delta^{(k,0)}\right]^c_L$.
Neglecting other contributions, the even stronger constraints
$\left[\Delta^{(1,0)}\right]_L^c\simeq 0.44 \left[\Delta^{(0,0)}\right]_L^c$
and $\left[\Delta^{(2,0)}\right]_L^c\simeq 
0.22 \left[\Delta^{(0,0)}\right]_L^c$ are obtained.
The $k=0,1,2$ $\left[ \Delta^{(k,0)}\right]^c_L$ implied
by the $O(\alpha_s^3)$ $D=2$ OPE truncation prescription (employing
a $k$-independent $m_s$) are, in contrast, in the ratios $1:1.16:1.42$,
badly violating even the weaker constraint,
Eq.~(\ref{rigorous}). 
Since the experimental spectral distribution necessarily respects spectral
positivity, independent fits for $m_s$ using different 
$(k,0)$ FESR's will, unavoidably, produce central
values having an unphysical decrease with $k$. Such a decrease 
is seen in all inclusive $(k,0)$ analyses.
A large portion of the observed instability-with-$k$ can
be attributed to the violation of spectral positivity~\cite{kmtau02}. 
Much improved stability is obtained for the
longitudinally-subtracted $(J)=(0+1)$ 
version of the $(k,0)$ analysis~\cite{pich040p1}. 

The absence of an experimental spin separation means that,
to avoid the above problems,
and work with the better-behaved $(J)=(0+1)$ sum rules,
one needs theoretical input for the unknown 
(continuum) part of the longitudinal spectral distribution.
The flavor $us$ A part can be obtained
from the results of Ref.~\cite{kmps01} (which determines
the excited $K$ decay constants from a sum rule analysis of
the flavor $us$ pseudoscalar correlator);
the flavor $us$ V part, similarly, from a
detailed study of the related flavor $us$ scalar correlator~\cite{jop}. 
Details may be found in the
original references{\begin{footnote}{These analyses can also be used to
obtain independent determinations of $m_s$; the consistency
of these determinations with those from the $(J)=(0+1)$ 
$\tau$ decay sum rules provides further support for their reliability.}
\end{footnote}}. It turns out that, even if one assigns very conservative
errors ($\sim 50\%$) to these determinations, the impact on
the uncertainties in the resulting longitudinally-subtracted $(J)=(0+1)$
sum rules is small. The reason is easily understood.

To simplify discussion, consider the narrow width approximation (NWA)
for the $K(1460)$ and $K_0^*(1430)$. The corresponding decay constants, $f$,
vanish in the $SU(3)$ chiral limit and hence
receive a chiral suppression (proportional to $m_s$) for physical $m_s$. The
corresponding longitudinal spectral contributions ($\propto f^2$) 
are thus doubly chirally suppressed relative to
the $\pi$ and $K$ pole terms. In the NWA, taking the $(0,0)$
spectral weight case to be specific, the integrated longitudinal
contribution of a scalar or pseudoscalar state of mass $M$ is
proportional to
\begin{equation}
{\frac{2M^2}{m_\tau^2}}\left( 1-{\frac{M^2}{m_\tau^2}}\right)^2\, f^2\ .
\label{longsubsmall}
\end{equation}
The kinematic factor, 
${\frac{2M^2}{m_\tau^2}}\left( 1-{\frac{M^2}{m_\tau^2}}\right)^2$,
is $0.13$ for the $K$ and $\sim 0.15$ for the
$K_0^*(1430)$ and $K(1460)$. Thus, the relative size of
the integrated longitudinal continuum and $K$ pole contributions
is determined almost entirely by the square of the ratio of 
the corresponding decay constants, and is doubly chirally suppressed.
Since the $K$ pole contribution is very accurately known,
even rather large errors on the continuum contribution 
will correspond to small errors on the full (pole plus continuum) 
longitudinal spectral integral. The subtraction needed to go from the 
inclusive experimental $(0,0)$ spectral integral 
to the analogous $(J)=(0+1)$ component
thereof can thus be performed with good accuracy.
Extra factors of $\left( 1-M^2/m_\tau^2 \right)$, present for the higher
$(k,0)$ sum rules, will further suppress continuum contributions 
relative to the leading $K$ pole term, making the longitudinal
subtraction even more reliable.

In summary, (i) the conclusion that only non-inclusive,
$(J)=(0+1)$ flavor-breaking FESR's should be employed in future 
seems unavoidable, in view of the severity of the problems 
with the corresponding inclusive FESR's;
(ii) the longitudinal subtraction needed for a determination of
the spectral integrals of the $(J)=(0+1)$ sum rules is
dominated by the well-known $K$ pole term, and can be 
performed with good accuracy, with the results of Refs.~\cite{kmps01,jop} 
to be used for the small flavor $us$ continuum contributions;
(iii) having accepted the necessity of performing a longitudinal 
subtraction, one of the major arguments in favor
of the use of the $(k,m)$ spectral
weights (the possibility of avoiding a spin separation of
the experimental spectral data) is no longer operative, and
one is free to explore alternate weight choices
which may improve the accuracy with which the OPE and/or spectral
integral sides of the resulting flavor-breaking $(J)=(0+1)$ sum rules
can be evaluated.

We will return to the latter point below.

\subsection{Convergence of the Integrated $(0+1)$ OPE Series}
The necessity of a longitudinal subtraction means that one 
must focus on sum rules for the flavor-breaking $(J)=(0+1)$
correlator. Experimental and theoretical uncertainties are
reduced by working with the difference of the V+A sums
for the flavor $ud$ and $us$ cases. Two points, related to the question of 
the convergence of the integrated OPE series, require discussion: 
the behavior of the integrated $D=2$ series, and the treatment of
contributions with $D>6$.

\subsubsection{The $D=2$ OPE Series}
In the $\overline{MS}$ scheme, 
the $D=2$ term in the OPE of the $(J)=(0+1)$ V+A $us$ correlator
is proportional to $\left[ m_s(Q^2)\right]^2/Q^2$
times the series~\cite{chettau04}
\begin{equation}
1+{\frac{7}{3}}a+19.58 a^2+202.3a^3
+(2200\pm 200)a^4+\cdots
\end{equation}
where $a=\alpha_s(Q^2)/\pi$ and the $O(a^4)$ coefficient is the PMS
estimate based on the $O(a^3)$ result reported by Chetyrkin
at this meeting~\cite{chettau04}. Since $a(m_\tau^2)\simeq 0.10$,
the convergence of the last few terms of the series
is actually very slow at the spacelike point 
on the circle $\vert s\vert =s_0$, even at the largest 
scale, $s_0=m_\tau^2$, allowed by kinematics. As one moves
along the circle toward the timelike point, however,
the logarithmic running of $\alpha_s$ causes
$\vert \alpha_s\left( Q^2\right)\vert$ to decrease,
improving the convergence of the correlator series. 
Different choices of FESR
weight, which emphasize different regions of the circle,
can thus lead to integrated $D=2$ series with significantly
different convergence behaviors. Within the $(k,0)$ spectral weight
family, e.g., one expects increasing $k$ to produce slower
convergence, since the additional factors of $(1-s/s_0)$ 
weight more and more heavily contributions from the part of
the circle near the spacelike point, where convergence is slowest. 
Cancellations on the contour
can also play a role in determining the convergence of the integrated
series.

The convergence behavior of the integrated $ud$-$us$ V+A, $(J)=(0+1)$
$D=2$ OPE series for the $(k,0)$ spectral weights is illustrated in Table
1. The contributions have been evaluated using CIPT, for $s_0=m_\tau^2$.
The results are normalized to the leading ($O(a^0)$) term.
We have included an $O(a^4)$ contribution generated using the PMS
estimate for the $O(a^4)$ correlator coefficient{\begin{footnote}{It 
is worth noting that the analogous estimate for the $O(a^3)$ 
coefficient~\cite{bck} turned out to be reliable with an accuracy 
of $\sim 1\%$.}\end{footnote}}.
The pattern of convergence is typically somewhat worse if 
one works with the Adler function, rather than the correlator
itself (see Chetyrkin's talk at this meeting for more on this
point). The convergence also deteriorates significantly
as $s_0$ decreases. Note that the convergence of the
$(0,0)$ series is {\it not} good, despite the impression 
an $O(a^2)$ truncation might give. The $O(a^2)$ contribution
happens to be small because of cancellations among contributions
from different parts of the OPE contour; this
cancellation, however, is ``accidental'', in the sense
that it does not persist for higher order contributions. Similar
accidental cancellations occur for the other $(k,0)$ cases, though
at orders which increase with $k$; as a result, only a hint of this
behavior shows up in the table for $k=2,3,4$. 

\begin{table}
\begin{center}
\caption{Convergence of integrated $(0+1)$ $D=2$ OPE series
for the $(k,0)$, $k=0,\cdots ,4$ spectral weights}
\label{table:1}
\vskip .15in\noindent
\begin{tabular}{@{}lrrrrr}
\hline
Weight:&(0,0)&(1,0)&(2,0)&(3,0)&(4,0)\\
\hline
$O(a^0)$&1&1&1&1&1\\
$O(a^1)$&.14&.21&.26&.30&.33\\
$O(a^2)$&-.01&.10&.19&.27&.34\\
$O(a^3)$&-.18&-.04&.09&.21&.34\\
$O(a^4)$&-.38&-.23&-.08&.09&.28\\
\hline
\end{tabular}\\[2pt]
\end{center}
\end{table}

The disappointing convergence of the $D=2$ series for
the $(k,0)$ spectral weights is not a general feature of
flavor-breaking V+A, $(J)=(0+1)$ FESR's. In fact, by studying
the behavior of the correlator in the complex plane,
it is possible to construct weights which emphasize precisely
those regions of the contour where the OPE is not only
reliable but displays improved $D=2$ convergence~\cite{kmms00}. 
Three weights of this type were discussed in Ref.~\cite{kmms00}.
For these weights, in contrast to the $(k,0)$ spectral weights,
the suppression of higher order integrated $D=2$ contributions results from
a dominance by the region of improved correlator
convergence, and not from an order-dependent accidental cancellation
along the contour. For this reason, the improved convergence persists 
even to much higher orders~\cite{kmms00}. 
Employing the information reported by Chetyrkin
for the values of the $O(a^3,a^4)$ coefficients~\cite{chettau04}, the 
$s_0=m_\tau^2$ contour-improved integrated $D=2$ series for
these weights behave as
\begin{eqnarray}
&&1+.26+.21+.17+.11\ ({\rm for\ }w_{20})\nonumber\\
&&1+.23+.17+.11+.05\ ({\rm for\ }w_{10}),\ {\rm and}\nonumber\\
&&1+.25+.19+.15+.10\ ({\rm for\ }\hat{w}_{10})\ .
\end{eqnarray}

In summary, 
(i) the convergence of the integrated $(J)=(0+1)$, V+A $ud$-$us$ 
$D=2$ series for the $(k,0)$ spectral weights 
is problematic (this is particularly true of
the $(0,0)$ case, although this fact does not become evident
until one goes beyond $O(a^2)$);
(ii) alternate weight choices exist with improved $D=2$ 
convergence.{\begin{footnote}{Additional tests of this improved convergence,
through comparison to the results obtained using, 
instead of the truncated correlator, the truncated Adler function,
may be found in the original conference talk; space constraints
preclude a discussion of this point here.}\end{footnote}}

\subsubsection{$D>6$ OPE Contributions}
The OPE series for the $ud$-$us$ $(J)=(0+1)$, V+A correlator difference is
known up to terms of dimension $D=6$. The $D=4$ contribution is
well determined phenomenologically, and the $D=6$ term can be estimated
using the vacuum saturation approximation (VSA). It turns out that
$D=6$ contributions to those FESR's studied in the literature are
small, even if one assigns a 
factor $5-10$ error to the VSA result{\begin{footnote}
{Such an error estimate should be considered extremely conservative
in view of the results for those V, A current correlator combinations
for which the VSA has been 
explicitly tested~\protect\cite{vectord6,cdgd6}.}\end{footnote}}.
$D>6$ contributions are not known; nor are phenomenological
values available for a full set of $D>6$ condensates.
In existing flavor-breaking $\tau$-decay-based analyses,
$D>6$ contributions have been {\it assumed} to be safely
negligible at the scales employed, usually without explicit
tests of this assumption. We point out below how
such tests may be carried out, and explain 
why, for certain of the $(k,0)$ spectral weights, results
obtained in the absence of such tests should be viewed
with caution.

Consider a polynomial weight, $w(y)=\sum_{k=0}^N b_k y^k$, written 
in terms of the natural variable $y=s/s_0$. The ``pinching'' condition
(necessary for the reliability of the OPE at intermediate
scales) is $w(1)=0$. A term $C_D/Q^D$
in the OPE yields a contribution to the $w(y)$-weighted OPE integral
proportional to $b_{\kappa_D}\, C_D/s_0^{\kappa_D}$,
with $\kappa_D=(D-2)/2$.
The fact that integrated OPE contributions with different $D$
scale differently with $s_0$ allows one to test
the assumption that higher $D$ terms
are safely negligible by studying
the $s_0$ dependence of any sum rule output. 

As an example,
the $(4,0)$ weight, having degree 7, in 
principle produces integrated OPE contributions up to
$D=16$. If a nominal determination of, e.g., $m_s$ using the $(4,0)$
FESR has incorrectly assumed that $D>6$ terms can be
neglected, this will show up as a variation of $m_s$ with $s_0$.
This variation results from 
the fact that the integrated $D=2$ term, from which  $m_s$
is determined, scales as a constant (up to logarithmic
corrections), but has been forced to absorb the effect of contributions
with $D>6$, which scale like $1/s_0^N$, with $N\geq 3$.
It is crucial to perform this $s_0$-stability test, especially 
for polynomials $w(y)$ having coefficients, $b_k$, with $k\geq 3$,
which are large.
 
For the $(k,0)$ spectral weights, the relevant
polynomial coefficients grow with $k$; hence so does
the danger of neglecting $D>6$ contributions.
The explicit forms of the $(J)=(0+1)$ $(k,0)$ weights, 
$w^{(k,0)}(y)=(1-y)^k\, w_T(y)$, are
\begin{eqnarray}
{w}^{(0,0)}(y)&=&1-3y^2+2y^3\nonumber\\
{w}^{(1,0)}(y)&=&1-y-3y^2+5y^3-2y^4\nonumber\\
{w}^{(2,0)}(y)&=&1-2y-2y^2+8y^3-7y^4+2y^5\nonumber\\
{w}^{(3,0)}(y)&=&1-3y+10y^3-15y^4+9y^5-2y^6\nonumber\\
{w}^{(4,0)}(y)&=&1-4y+3y^2+10y^3-25y^4+24y^5\nonumber\\
&&\ \ \ \ \ \-11y^6+2y^7\ .
\end{eqnarray}
The $O(y^4)$ and $O(y^5)$ coefficients in $w^{(4,0)}(y)$, e.g.,
which govern the integrated $D=10$ and $12$ OPE contributions 
to the $(4,0)$ FESR, are
more than an order of magnitude larger than the $O(y^3)$ coefficient
in $w^{(0,0)}(y)$, which governs the integrated $D=8$ contribution
to the $(0,0)$ FESR. Neglect of $D>6$
contributions is thus far safer in the $(0,0)$ than in the $(4,0)$ case.

The alternate weights of Ref.~\cite{kmms00} were
constructed to have coefficients $b_k$, $k\geq 3$,
as small as possible, given other constraints.
With the $O(y^0)$ terms normalized to $1$, as for the $(k,0)$
weights, the largest of these $b_k$'s is
1.206 for $\hat{w}_{10}$, $2$ for $w_{10}$ and $2.087$ for $w_{20}$.
Neglect of $D>6$ contributions is thus {\it much}
safer than it is for the higher $(k,0)$ spectral weights, though one
should, of course, still perform $s_0$-stability checks in all cases.


\subsection{Cancellations in the
$ud$-$us$ Spectral Difference}
One might naively expect the cancellation on the spectral integral
side of a flavor-breaking $ud$-$us$ FESR to be at the
$\sim 20-30\%$ level, the typical scale of $SU(3)_F$ breaking.
With present $ud$-$us$ spectral integral errors dominated by
the $us$ contribution, and these errors being at the $\sim 3\%$ 
level for $s_0=m_\tau^2$, $\sim 10\%$ errors would then be
expected for the $ud$-$us$ difference.
Unfortunately, this naive estimate is not borne out: for
weights considered in the literature, the
cancellation is much stronger, leading to much
larger fractional errors on the $ud$-$us$ difference.
Since the level of cancellation depends on the weight,
$w(y)$, the accuracy of the extraction of a quantity such as $m_s$, 
from a given set of experimental data, can be improved 
by a judicious choice of weight(s).

Table~\ref{table2} shows the ratio of $ud$-$us$ to $ud$
spectral integrals for $s_0=m_\tau^2$, for
$(J)=(0+1)$, V+A FESR's based on
(i) the $(k,0)$ spectral weights, and 
(ii) the weights, $w_{10}$, $\hat{w}_{10}$, and
$w_{20}$ of Ref.~\cite{kmms00}. 
CKMU labels results corresponding to the central 
PDG04 unitarity-constrained fit values,
$V_{ud}=0.9745$, $V_{us}=0.2240$, CKMN
results corresponding to the central PDG04 independent fit values,
$V_{ud}=0.9738$, $V_{us}=0.2200$. 

\begin{table}
\begin{center}
\caption{$ud$-$us$ spectral cancellation level for various
weights and for two choices of the input CKM values, $V_{ud}$
and $V_{us}$}
\vskip .05in
\label{table2}
\begin{tabular}{lrr}
\hline
Weight&CKMN&CKMU\\
\hline
$(0,0)$&0.6$\%$&4.0$\%$\\
$(1,0)$&4.5$\%$&7.8$\%$\\
$(2,0)$&8.3$\%$&11.4$\%$\\
$(3,0)$&11.9$\%$&14.9$\%$\\
$(4,0)$&15.6$\%$&18.5$\%$\\
$w_{10}$&3.7$\%$&7.0$\%$\\
$\hat{w}_{10}$&4.4$\%$&7.7$\%$\\
$w_{20}$&7.5$\%$&10.6$\%$\\
\hline
\end{tabular}
\end{center}
\end{table}

\begin{figure}
\caption{Stability of various FESR analyses for $V_{us}$. The vertical
axis shows $V_{us}$, the horizontal axis $s_0$ in GeV$^2$. 
The curves, from top to bottom on the LHS of the figure, correspond to the
weights $w^{(4,0)}$, $w^{(3,0)}$, $w_{20}$, $w^{(2,0)}$ (dashed line),
$w_{10}$ (solid line) and $w^{(0,0)}$.}
\includegraphics[height=2.8in,width=3.8in,
angle=270]{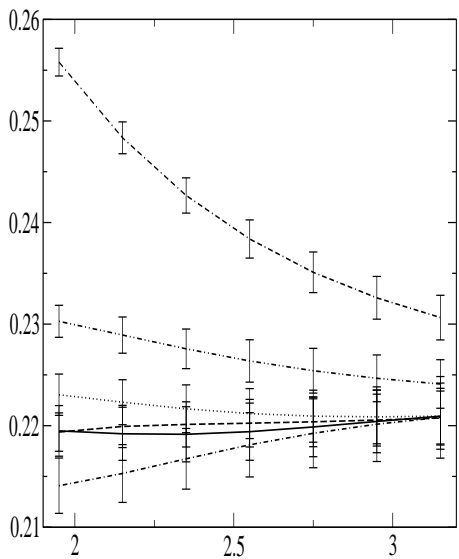}
\label{vusstability}
\end{figure}

The results show a high level of sensitivity to CKM input
for those weights having the strongest cancellation. The
strong cancellation also leads to large fractional
errors for the integrated $ud$-$us$ differences.
This effect is responsible, e.g., for the large errors quoted 
on $(0,0)$ spectral weight FESR determinations of $m_s$
in the literature. Large shifts in the 
central values of $m_s$ caused by apparently rather small
changes in the total strange branching fraction value are also the result
of this high level of cancellation. (Another
example of this effect will be seen in the next section.) The cancellation
is at a much more acceptable level for the $(3,0)$ and
$(4,0)$ weights. Unfortunately, as we have seen
above, these weights have rather slow
$D=2$ convergence, as well as large coefficients
which make neglect of $D>6$ contributions more problematic.
More accurate $us$ data, especially in the region above
$\sim 1\ {\rm GeV}^2$, might allow alternate weights
to be constructed which deal with this problem more effectively,
without producing a deterioration in the $ud$-$us$
error situation.

In view of the above-noted ``close cancellations'', it is
desirable to reduce, where feasible,
the error on any particular $ud$-$us$ spectral integral. 
Evaluating the $\pi$ and $K$ pole term contributions
using the more accurate $f_\pi V_{ud}$ and $f_K V_{us}$ 
$\pi_{\ell 2}$ and $K_{\ell 2}$ determinations is helpful in this regard.
While one should obviously check that
$\tau$ decay determinations of the $\tau\rightarrow\pi\nu_\tau$
and $\tau\rightarrow K\nu_\tau$ BR's are compatible with Standard
Model (SM) expectations based on $\pi_{\ell 2}$ and $K_{\ell 2}$
data, attempts to extract $m_s$ and/or $V_{us}$ from $\tau$
decay data are predicated on the assumption that beyond-the-SM
effects may be neglected in $\tau$ decay; there is, thus,
no compelling reason to use the larger-error $\tau$-decay-based
pole term values in such analyses.

\section{An Illustrative Example}
We illustrate some features of the above discussion by
considering a determination of $V_{us}$ analogous to
that reported in Ref.~\cite{pich040p1}. A value 
$m_s(2 \ {\rm GeV}^2) =95\pm 20$ MeV, representing an average 
(with conservative errors) from non-$\tau$-decay based determinations
is used on the OPE side of the various FESR's. The corresponding
spectral integrals are based on the recently-reported OPAL update of the
$us$ spectral database~\cite{opal04}, with the following
caveats. At present, neither the numerical values of the OPAL
$us$ spectral distribution nor the corresponding
covariance matrix have been made publicly available. OPAL has reported spectral
integrals, with fully correlated errors, 
only for the $(k,0)$ spectral weights, 
and, for these, only at $s_0=m_\tau^2$. To work out spectral 
integrals for either non-spectral weights, or for spectral
weights, but at $s_0<m_\tau^2$, we follow the strategy used
previously by ALEPH (see the two papers by S. Chen et al.
in Ref.~\cite{mstauinclusive}). Explicitly, we start with the 1999
ALEPH $us$ spectral distribution~\cite{aleph99} and rescale
that part of the spectral distribution associated with each exclusive
mode by the corresponding ratio of current to ALEPH 1999
branching fractions. To guide the eye we include ``experimental''
errors generated using the publicly-available ALEPH covariance
matrix. We have also employed $\pi_{\ell 2}$ and $K_{\ell 2}$
results in evaluating the $\pi$ and $K$ pole term contributions.
The results of this exercise for a selection of the weights
discussed above, as a function of $s_0$, are shown in
Fig.~\ref{vusstability}. To avoid (further) cluttering the figure,
OPE errors have not been included.

The figure shows considerable instability for the
$(0,0)$, $(3,0)$ and $(4,0)$ spectral weights. Good stability
is observed for $w_{10}$, $w_{20}$ and the $(2,0)$ spectral weight.
The instability of the $(0,0)$ determination is almost certainly a consequence
of the poor convergence behavior of the integrated $D=2$ OPE
series. The $s_0$ dependence, together with the comparison of
the results for the range of different weights shown, gives one 
confidence in an extracted value of $V_{us}$
represented  by the convergence of the four lowest weight cases 
as $s_0\rightarrow m_\tau^2$ in the figure, $V_{us}\simeq 0.2209$.
More detailed results, with realistic experimental and OPE
errors will be reported elsewhere.

We conclude with a further illustration of the sensitivities of output
parameters to small changes in experimental data, resulting
from strong cancellations in the $ud$-$us$ spectral integral differences. 
The above result for $V_{us}$ employed the world average, $B=0.0033$,
for the $K^-\pi^+\pi^-$ branching fraction. If one instead shifts
to the average, $B=0.0040$, of the OPAL and CLEO determinations,
which are in good agreement, and not in good agreement with ALEPH,
the central value of $V_{us}$, e.g., 
from the good-stability $w_{10}$ extraction,
changes from $0.2209$ to $0.2232$.

Further details of the $V_{us}$ analysis, and of the related
$m_s$ analysis, will be reported elsewhere.

\end{document}